\documentclass[10pt,final]{IEEEtran}

\usepackage[utf8]{inputenc} 
\usepackage[T1]{fontenc}    
\usepackage{url}            
\usepackage{booktabs}       
\usepackage{amsfonts}       
\usepackage{nicefrac}       
\usepackage{microtype}      

\usepackage{algpseudocode}
\usepackage{algorithm}
\usepackage{comment}
\usepackage{tikz}
\usepackage{graphicx}
\usepackage{pgfplots}
\pgfplotsset{compat=1.14}
\usetikzlibrary{positioning, fit, arrows.meta, shapes}
\newcommand{\empt}[2]{$#1^{\langle #2 \rangle}$}

\algblock{ParFor}{EndParFor}
\algnewcommand\algorithmicparfor{\textbf{parfor}}
\algnewcommand\algorithmicpardo{\textbf{do}}
\algnewcommand\algorithmicendparfor{\textbf{end\ parfor}}
\algrenewtext{ParFor}[1]{\algorithmicparfor\ #1\ \algorithmicpardo}
\algrenewtext{EndParFor}{\algorithmicendparfor}

\usepackage{graphicx}
\usepackage{subfigure}

\usepackage{amsmath}
\usepackage{amssymb}
\usepackage{mathtools}

\usepackage{amsfonts}
\usepackage{amsmath,amsfonts,amssymb,amsthm,epsfig,epstopdf,titling,url,array}

\theoremstyle{plain}

\theoremstyle{definition}

\theoremstyle{remark}

\usepackage{tikz}

\IEEEoverridecommandlockouts

\begin{document}

\title{
Designing knowledge plane to optimize leaf and spine data center }

\author{%
Mujahid Sultan$^{\dag}$, Dodi Imbuido$^{\ddag}$, Kam Patel$^{\ddag}$, James MacDonald$^{\ddag}$ and Kumar Ratnam$^{\ddag}$%
 \\ $^{\dag}$Department of Computer Science, Ryerson University, Toronto, Canada\\
 $^{\ddag}$Aytra Inc. 18 Wynford Dr. Toronto, Canada\\
 mujahid.sultan@ryerson.ca; \{dodi.imbuido, james.macdonald, kam.patel, kumar\}@aytra.com
 ~\thanks{Note: A slightly modified version of this pre-print is accepted for publication by IEEE Cloud 2020.}
}

\date{}

\maketitle

\begin{abstract}
In the last few decades, data center architecture evolved from the traditional client-server to access-aggregation-core architectures. Recently there is a new shift in the data center architecture due to the increasing need for low latency and high throughput between server-to-server communications, load balancing and, loop-free environment. This new architecture, known as leaf and spine architecture, provides low latency and minimum packet loss by enabling the addition and deletion of network nodes on demand. 
Network nodes can be added or deleted from the network based on network statistics like link speed, packet loss, latency, and throughput.

With the maturity of Open Virtual Switch (OvS) and OpenFlow based Software Defined Network (SDN) controllers, network automation through programmatic extensions has become possible based on network statistics. Separation of \textit{control plane} and \textit{data plane} has enabled automated management of network and Machine Learning (ML) can be applied to learn and optimize the network. 

In this publication, we propose the design of an ML-based approach to gather network statistics and build a \textit{knowledge plane}. We demonstrate that this knowledge plane enables data center optimization using southbound APIs and SDN controllers. We describe the design components of this approach - using a network simulator and show that it can maintain the historical patterns of network statistics to predict future growth or decline. We also provide an open-source software that can be utilized in a leaf and spine data center to provide elastic capacity based on load forecasts.



\end{abstract}

\section{Introduction}

\subsection{Background}

The traditional design of IP networks is decentralized - for resiliency. Networking devices (e.g., switches and routers) are designed to be independent - by coupling the routing and the data. Performance is mainly achieved by merchant silicon or custom application-specific integrated circuits, thus leading to inflexible control of the devices. Recently with the ONF~\cite{specifiationv1} consortium's push to produce networking devices equipped with south-bound APIs (OpenFlow~\cite{mckeown2008openflow}) the networking devices have become more accessible. Traditional switches used to have a single unit responsible for forwarding policy and a physical layer which carried out these policies. With the advent of OvS~\cite{onf20121}, OpenFlow and SDN controllers~\cite{kirkpatrick2013software}, the physical layer called \textit{data plane} is separated from the policy layer called \textit{control plane} - opening new opportunities for automation.

Maturity of Open Virtual Switch (OvS) has led to the development of many enterprise-class SDN controllers also known as network operating systems (NOS), for example, NOX~\cite{gude2008nox}, ONOS~\cite{berde2014onos}, OpenDaylight~\cite{medved2014opendaylight} and Ryu~\cite{morita2012ryu}. These SDN controllers separate policies from the data. 

The \textit{data plane}, responsible for forwarding packets, relies on the \textit{control plane} for the forwarding rules or policies. The \textit{control plane} keeps and manages policies, which are being called \textit{flow rules}, traditionally known as forwarding rules. These \textit{flow rules} dictate the networking policies to the \textit{data plane}. The \textit{data plane} is kept unaware of the network topology and relies on the \textit{control plane} to populate their forwarding tables. Though this is a fairly new paradigm, however, it has seen success in many business domains~\cite{feamster2013road}, in the form of policies managed and deployed to hundreds of network devices from a single centralized \textit{control plane}. 

When the policies are developed independently of the historical network traffic patterns, these are called \textit{static policies}. If the policies are learned based on the historical network traffic, these are called \textit{dynamic policies}. In this publication, we stream network performance metrics to Machine Learning (ML) layer and build \textit{dynamic policies} based on predictions by the neural network. We use these \textit{dynamic policies} to add and or remove spine nodes to achieve desired network topologies in the leaf and spine data center.



\subsection{Motivation}

Dynamic policies have been applied to the network devices using numerous network optimization and improvement tasks such as network availability \cite{nencioni2017impact}, MPLS optimization \cite{husni2018design}, load balancing in data centers \cite{kim2017load}, and resilience for smart grid communications \cite{aydeger2015sdn}. To date, there has not been any significant development that uses ML in developing the \textit{dynamic policies}. 

In this study, we created a spine and leaf network and learned network traffic patterns using ML. Based on patterns found in the ML layer, we update flow rules in real-time using southbound APIs to optimize the network.

\textbf{Road map:} The rest of the paper is organized as follows: Related work is summarized in Section~\ref{relatedwork}. In Section~\ref{methods} we describe experimental setup for network simulation, control plane and the forecast engine. In Section~\ref{results} we discuss the results and in Section~\ref{challenges} we present some future directions.

\section{Literature Review} \label{relatedwork}

A comprehensive list of SDN controllers and their performance and maturity comparison is given by \cite{salman2016sdn, kreutz2014software}, and an overview of knowledge defined networking can be found at \cite{mestres2017knowledge}. Monitoring latency with OpenFlow is investigated by \cite{phemius2013monitoring}. Data path performance of the spine-leaf data center is described by \cite{alizadeh2013data}. An OpenDaylight based MPLs controller is investigated by \cite{husni2018design}. To improve the quality of service, \cite{okafor2017leveraging} used leaf and spine architecture. Long short-term memory (LSTM) \cite{hochreiter1997long} based traffic prediction method to assist light path reconfiguration in hybrid data center networks is investigated by \cite{shi2018lstm}. 
None of these studies investigated network optimization using LSTM from the entire metric set of the leaf and spine data center perspective nor implemented any streaming engine.

\section{Experimental Setup}\label{methods}
\begin{figure}[!htb]
\centering
    \includegraphics[height=7.85cm, width = .80\columnwidth]{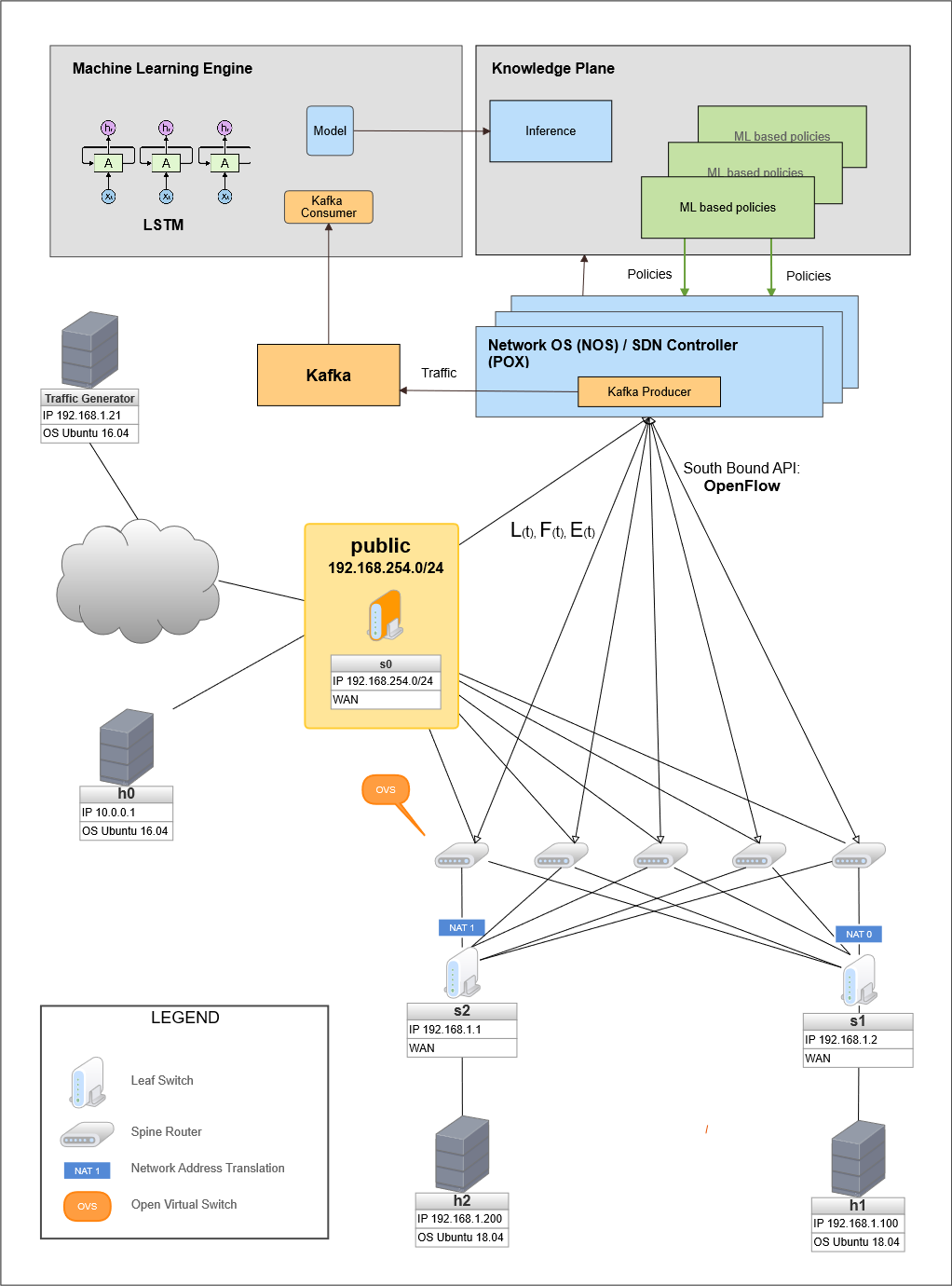}
    \caption{\label{fig:topology} The Architecture of a leaf and spine network with two "leaf" and five "spine" nodes. We use the POX SDN controller with southbound OpenFlow based APIs to access the network. Network traffic is streamed to the ML layer using Kafka topics. LSTM neural network is used to interpret and forecast network metrics. The inference is passed to the Knowledge Plane which controls the network in real-time (by adding or deleting nodes).}
\end{figure}

Monitoring latency with OpenFlow was investigated by~\cite{phemius2013monitoring}. We use this method and software in our experiments. To conduct our experiments, we simulated a leaf and spine network using Mininet~\cite{handigol2012reproducible}, as shown in Fig.~\ref{fig:topology}. We streamed the spine switch layer traffic to a Kafka \cite{kreps2011kafka} server. In the ML layer, we used the LSTM neural network-based application, which consumes Kafka streams to predict spine node latency. We are developing knowledge plane components to deploy \textit{dynamic policies} to the network via southbound APIs. The design components of our application and the data center simulation are briefly described below.

\subsection{The network}
Mininet\footnote{Available at https://github.com/mininet} is a network emulator to mimic large networks on a single computer or virtual machine developed by \cite{handigol2012reproducible}. We created a leaf and spine network segment with five spine and three leaf nodes using Mininet and transferred some large files between leaf nodes to generate the traffic. 

\subsection{The SDN Controller (POX)}
We used POX\footnote{available at: https://github.com/noxrepo/pox}, a python implementation of NOX~\cite{gude2008nox} SDN controller which is based on the OvS~\cite{pfaff2015design}. A python script integrates POX to capture and stream network traffic. We used the Kafka-python\footnote{https://kafka-python.readthedocs.io/en/master/} to generate traffic topics, which are consumed by the ML layer. 

\subsection{The Forecast Engine }
\label{sec:lstm}
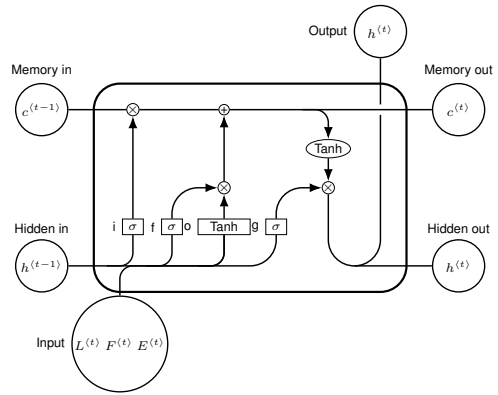
\begin{figure}[!thb]
\centering 
\resizebox{.75\columnwidth}{!}{%
\begin{tikzpicture}[
    font=\sf \scriptsize,
    >=LaTeX,
    cell/.style={
        rectangle, 
        rounded corners=5mm, 
        draw,
        very thick,
        },
    operator/.style={
        circle,
        draw,
        inner sep=-0.5pt,
        minimum height =.2cm,
        },
    function/.style={
        ellipse,
        draw,
        inner sep=1pt
        },
    ct/.style={
        circle,
        draw,
        line width = .75pt,
        minimum width=1cm,
        inner sep=1pt,
        },
    gt/.style={
        rectangle,
        draw,
        minimum width=4mm,
        minimum height=3mm,
        inner sep=1pt
        },
    mylabel/.style={
        font=\scriptsize\sffamily
        },
    ArrowC1/.style={
        rounded corners=.25cm,
        thick,
        },
    ArrowC2/.style={
        rounded corners=.5cm,
        thick,
        },
    ]
    \node [cell, minimum height =4cm, minimum width=6cm] at (0,0){} ;

    \node [gt, label={[mylabel]left:i}] (ibox1) at (-2.25,-0.75) {$\sigma$};
    \node [gt, label={[mylabel]left:f}] (ibox2) at (-1.5,-0.75) {$\sigma$};
    \node [gt, minimum width=1cm, label={[mylabel]left:o}] (ibox3) at (-0.5,-0.75) {Tanh};
    \node [gt,, label={[mylabel]left:g}] (ibox4) at (0.5,-0.75) {$\sigma$};

    \node [operator] (mux1) at (-2.25,1.5) {$\times$};
    \node [operator] (add1) at (-0.5,1.5) {+};
    \node [operator] (mux2) at (-0.5,0) {$\times$};
    \node [operator] (mux3) at (1.5,0) {$\times$};
    \node [function] (func1) at (1.5,0.75) {Tanh};

    \node[ct, label={[mylabel]Memory in}] (c) at (-4,1.5) {\empt{c}{t-1}};
    \node[ct, label={[mylabel]Hidden in}] (h) at (-4,-1.5) {\empt{h}{t-1}};
    \node[ct, label={[mylabel]left:Input}] (x) at (-2.5,-3) {\empt{L}{t} \empt{F}{t} \empt{E}{t}};

    \node[ct, label={[mylabel]Memory out}] (c2) at (4,1.5) {\empt{c}{t}};
    \node[ct, label={[mylabel]Hidden out}] (h2) at (4,-1.5) {\empt{h}{t}};
    \node[ct, label={[mylabel]left:Output}] (x2) at (2.5,3) {\empt{h}{t}};

    \draw [ArrowC1] (c) -- (mux1) -- (add1) -- (c2);

    \draw [ArrowC2] (h) -| (ibox4);
    \draw [ArrowC1] (h -| ibox1)++(-0.5,0) -| (ibox1); 
    \draw [ArrowC1] (h -| ibox2)++(-0.5,0) -| (ibox2);
    \draw [ArrowC1] (h -| ibox3)++(-0.5,0) -| (ibox3);
    \draw [ArrowC1] (x) -- (x |- h)-| (ibox3);

    \draw [->, ArrowC2] (ibox1) -- (mux1);
    \draw [->, ArrowC2] (ibox2) |- (mux2);
    \draw [->, ArrowC2] (ibox3) -- (mux2);
    \draw [->, ArrowC2] (ibox4) |- (mux3);
    \draw [->, ArrowC2] (mux2) -- (add1);
    \draw [->, ArrowC1] (add1 -| func1)++(-0.5,0) -| (func1);
    \draw [->, ArrowC2] (func1) -- (mux3);

    \draw [-, ArrowC2] (mux3) |- (h2);
    \draw (c2 -| x2) ++(0,-0.1) coordinate (i1);
    \draw [-, ArrowC2] (h2 -| x2)++(-0.5,0) -| (i1);
    \draw [-, ArrowC2] (i1)++(0,0.2) -- (x2);

\end{tikzpicture}
}
    \caption{\label{fig:lstm} Design of a single LSTM unit with the input of three network metrics ($L_t$, $E_t$, and $F_t$) streamed from the python-Kafka engine.}
\end{figure}

The \textit{Kafka-python consumer} is used to gather network traffic streams. We used the LSTM neural network for predicting the network metrics. Fig.~\ref{fig:lstm} shows a single layer of the typical LSTM. We experimented with several architectures of LSTM to lower the validation error for the network latency. Below we briefly describe the input of the LSTM for our network simulation.

The Latency of a \textit{link} at time $t$ is defined by vector $L_t$, the \textit{fabric link speed} by vector $F_t$ and edge or \textit{leaf link speed} by vector $E_t$ giving us time series for each link in the spine network. These metrics can be written in vector form as:\\
$L_t = 
\begin{bmatrix}
L_{t-n},L_{t-(n-1)}, \dots ,L_{t-2},L_{t-1}
\end{bmatrix}$,\\ 
$F_t = 
\begin{bmatrix}
F_{t-n},F_{t-(n-1)}, \dots ,F_{t-2},F_{t-1}
\end{bmatrix}$ and \\
$ E_t = 
\begin{bmatrix}
E_{t-n},E_{t-(n-1)}, \dots ,E_{t-2},E_{t-1}
\end{bmatrix}
$.
And in matrix form for each spine \textit{link} can be written as:
\[
\begin{bmatrix}
L_t \\ 
F_t \\
E_t
\end{bmatrix} =
\begin{bmatrix}
L_{t-n},L_{t-(n-1)}, \dots ,L_{t-2},L_{t-1} \\
F_{t-n},F_{t-(n-1)}, \dots ,F_{t-2},F_{t-1}\\
E_{t-n},E_{t-(n-1)}, \dots ,E_{t-2},E_{t-1}
\end{bmatrix}
\]
A three-dimensional tensor represents the number of links with the third dimension for the switch label. Each input is normalized separately, and a different loss function is used, as the link latency is measured in different units than the speed. We implemented a 1-D convolutional net layer to sample the network traffic. Different architectures were tried. We used two-layered stacked LSTM with a random-dropout layer designed as three inputs and one output, as shown in Fig.~\ref{fig:lstm}. To train the model we used Keras~\cite{chollet2015keras} v2.3.0 python library with TensorFlow \cite{abadi2016tensorflow} v2.1 backend.

\section{Results and usage}\label{results}
\begin{figure}[!tb]
\centering
    \includegraphics[width = .84\columnwidth]{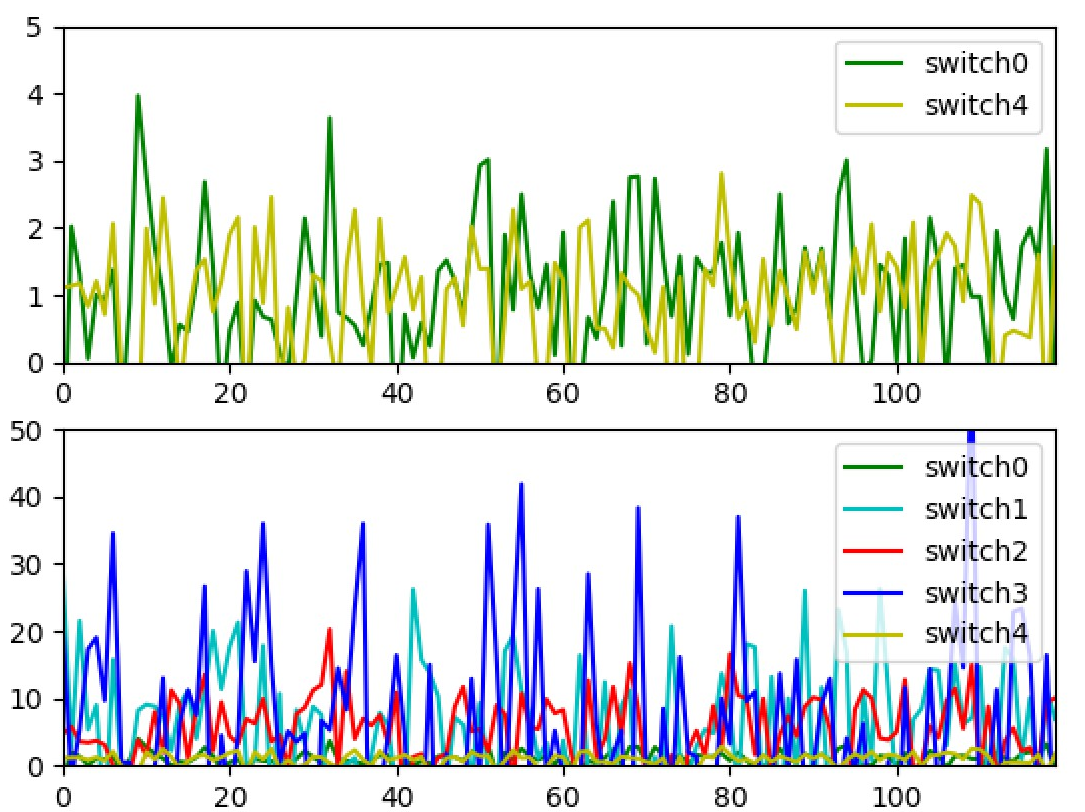}
    \caption{\label{fig:results} Top - Switches with low predicted latency; Bottom - Prediction of hourly latency by LSTM for all switches for 120 hours (microseconds). Switches with low predicted latency (switch 0 and 4) can be removed from the network without any impact on the overall health of the network.}
\end{figure}

Fig.~\ref{fig:results} (bottom part) shows the predicted average latency of the entire network for 120 hours (5 days). In the top part of Fig.~\ref{fig:results}, two low latency switches are shown, which are good candidates to be programmatically removed by the SDN controller - to save cost without any significant impact on the network health (switches below 6 microseconds latency). A similar analysis can be done if the overall predicted network latency goes beyond some predefined thresh-hold to add new nodes in the network. This programmatic addition and deletion of the network resources can make the leaf and spine networks cost-effective. 

In recent years, companies are opting for a more disaggregated approach to network equipment and software. The decoupling of the hardware and software has enabled companies to implement the best of the breed hardware with the most suitable software stack to avoid the common problem of vendor lock-ins. This decoupling has become the basis of improving architecture agility. The architecture and design of the ML-based control plane, described in this publication, enables companies to achieve the desired architectures.

\section{Code availability and future Research}\label{challenges}
Python code for data center simulation on Mininet using OvS, POX SDN controller, Kafka streaming engine, and LSTM layer can be found   at https://github.com/SultanMu/leaf-spine-knowledge-plane.git. 
We are currently developing south-bound SDN components to enforce dynamic policies to the network. This work has also opened other opportunities, with the availability of vXlan~\cite{mahalingam2014virtual}, we plan to learn and predict \textit{the edge}~\cite{shi2016edge} workloads and add/remove nodes (eastbound or westbound) in real-time. 
\newcommand{\BIBdecl}{\setlength{\itemsep}{-2pt}}
\bibliographystyle{IEEEtran}
\bibliography{IEEEabrv,main.bib}

\begin{thebibliography}{10}
\providecommand{\url}[1]{#1}
\csname url@samestyle\endcsname
\providecommand{\newblock}{\relax}
\providecommand{\bibinfo}[2]{#2}
\providecommand{\BIBentrySTDinterwordspacing}{\spaceskip=0pt\relax}
\providecommand{\BIBentryALTinterwordstretchfactor}{4}
\providecommand{\BIBentryALTinterwordspacing}{\spaceskip=\fontdimen2\font plus
\BIBentryALTinterwordstretchfactor\fontdimen3\font minus
  \fontdimen4\font\relax}
\providecommand{\BIBforeignlanguage}[2]{{%
\expandafter\ifx\csname l@#1\endcsname\relax
\typeout{** WARNING: IEEEtran.bst: No hyphenation pattern has been}%
\typeout{** loaded for the language `#1'. Using the pattern for}%
\typeout{** the default language instead.}%
\else
\language=\csname l@#1\endcsname
\fi
#2}}
\providecommand{\BIBdecl}{\relax}
\BIBdecl

\bibitem{specifiationv1}
O.~S. Specifiation, ``v1. 5, open network foundation, september 27, 2013.''

\bibitem{mckeown2008openflow}
N.~McKeown, T.~Anderson, H.~Balakrishnan, G.~Parulkar, L.~Peterson, J.~Rexford,
  S.~Shenker, and J.~Turner, ``Openflow: enabling innovation in campus
  networks,'' \emph{ACM SIGCOMM Computer Communication Review}, vol.~38, no.~2,
  pp. 69--74, 2008.

\bibitem{onf20121}
O.~S.~S. ONF, ``1.3. 0,'' 2012.

\bibitem{kirkpatrick2013software}
K.~Kirkpatrick, ``Software-defined networking,'' \emph{Communications of the
  ACM}, vol.~56, no.~9, pp. 16--19, 2013.

\bibitem{gude2008nox}
N.~Gude, T.~Koponen, J.~Pettit, B.~Pfaff, M.~Casado, N.~McKeown, and
  S.~Shenker, ``Nox: towards an operating system for networks,'' \emph{ACM
  SIGCOMM Computer Communication Review}, vol.~38, no.~3, pp. 105--110, 2008.

\bibitem{berde2014onos}
P.~Berde, M.~Gerola, J.~Hart, Y.~Higuchi, M.~Kobayashi, T.~Koide, B.~Lantz,
  B.~O'Connor, P.~Radoslavov, W.~Snow \emph{et~al.}, ``Onos: towards an open,
  distributed sdn os,'' in \emph{Proceedings of the third workshop on Hot
  topics in software defined networking}, 2014, pp. 1--6.

\bibitem{medved2014opendaylight}
J.~Medved, R.~Varga, A.~Tkacik, and K.~Gray, ``Opendaylight: Towards a
  model-driven sdn controller architecture,'' in \emph{Proceeding of IEEE
  International Symposium on a World of Wireless, Mobile and Multimedia
  Networks 2014}.\hskip 1em plus 0.5em minus 0.4em\relax IEEE, 2014, pp. 1--6.

\bibitem{morita2012ryu}
K.~Morita, I.~Yamahata, and V.~Linux, ``Ryu: Network operating system,'' in
  \emph{OpenStack Design Summit \& Conference}, 2012.

\bibitem{feamster2013road}
N.~Feamster, J.~Rexford, and E.~Zegura, ``The road to sdn,'' \emph{Queue},
  vol.~11, no.~12, pp. 20--40, 2013.

\bibitem{nencioni2017impact}
G.~Nencioni, B.~E. Helvik, A.~J. Gonzalez, P.~E. Heegaard, and A.~Kamisinski,
  ``Impact of sdn controllers deployment on network availability,'' \emph{arXiv
  preprint arXiv:1703.05595}, 2017.

\bibitem{husni2018design}
E.~Husni and A.~Bramantyo, ``Design and implementation of mpls sdn controller
  application based on opendaylight,'' in \emph{2018 International Symposium on
  Networks, Computers and Communications (ISNCC)}.\hskip 1em plus 0.5em minus
  0.4em\relax IEEE, 2018, pp. 1--5.

\bibitem{kim2017load}
T.~Kim, S.-G. Choi, J.~Myung, and C.-G. Lim, ``Load balancing on distributed
  datastore in opendaylight sdn controller cluster,'' in \emph{2017 IEEE
  Conference on Network Softwarization (NetSoft)}.\hskip 1em plus 0.5em minus
  0.4em\relax IEEE, 2017, pp. 1--3.

\bibitem{aydeger2015sdn}
A.~Aydeger, K.~Akkaya, and A.~S. Uluagac, ``Sdn-based resilience for smart grid
  communications,'' in \emph{2015 IEEE Conference on Network Function
  Virtualization and Software Defined Network (NFV-SDN)}.\hskip 1em plus 0.5em
  minus 0.4em\relax IEEE, 2015, pp. 31--33.

\bibitem{salman2016sdn}
O.~Salman, I.~H. Elhajj, A.~Kayssi, and A.~Chehab, ``Sdn controllers: A
  comparative study,'' in \emph{2016 18th Mediterranean Electrotechnical
  Conference (MELECON)}.\hskip 1em plus 0.5em minus 0.4em\relax IEEE, 2016, pp.
  1--6.

\bibitem{kreutz2014software}
D.~Kreutz, F.~M. Ramos, P.~E. Verissimo, C.~E. Rothenberg, S.~Azodolmolky, and
  S.~Uhlig, ``Software-defined networking: A comprehensive survey,''
  \emph{Proceedings of the IEEE}, vol. 103, no.~1, pp. 14--76, 2014.

\bibitem{mestres2017knowledge}
A.~Mestres, A.~Rodriguez-Natal, J.~Carner, P.~Barlet-Ros, E.~Alarc{\'o}n,
  M.~Sol{\'e}, V.~Munt{\'e}s-Mulero, D.~Meyer, S.~Barkai, M.~J. Hibbett
  \emph{et~al.}, ``Knowledge-defined networking,'' \emph{ACM SIGCOMM Computer
  Communication Review}, vol.~47, no.~3, pp. 2--10, 2017.

\bibitem{phemius2013monitoring}
K.~Phemius and M.~Bouet, ``Monitoring latency with openflow,'' in
  \emph{Proceedings of the 9th International Conference on Network and Service
  Management (CNSM 2013)}.\hskip 1em plus 0.5em minus 0.4em\relax IEEE, 2013,
  pp. 122--125.

\bibitem{alizadeh2013data}
M.~Alizadeh and T.~Edsall, ``On the data path performance of leaf-spine
  datacenter fabrics,'' in \emph{2013 IEEE 21st annual symposium on
  high-performance interconnects}.\hskip 1em plus 0.5em minus 0.4em\relax IEEE,
  2013, pp. 71--74.

\bibitem{okafor2017leveraging}
K.~C. Okafor, I.~E. Achumba, G.~A. Chukwudebe, and G.~C. Ononiwu, ``Leveraging
  fog computing for scalable iot datacenter using spine-leaf network
  topology,'' \emph{Journal of Electrical and Computer Engineering}, vol. 2017,
  2017.

\bibitem{hochreiter1997long}
S.~Hochreiter and J.~Schmidhuber, ``Long short-term memory,'' \emph{Neural
  computation}, vol.~9, no.~8, pp. 1735--1780, 1997.

\bibitem{shi2018lstm}
H.~Shi and C.~Wang, ``Lstm-based traffic prediction in support of periodically
  light path reconfiguration in hybrid data center network,'' in \emph{2018
  IEEE 4th International Conference on Computer and Communications
  (ICCC)}.\hskip 1em plus 0.5em minus 0.4em\relax IEEE, 2018, pp. 1124--1128.

\bibitem{handigol2012reproducible}
N.~Handigol, B.~Heller, V.~Jeyakumar, B.~Lantz, and N.~McKeown, ``Reproducible
  network experiments using container-based emulation,'' in \emph{Proceedings
  of the 8th international conference on Emerging networking experiments and
  technologies}, 2012, pp. 253--264.

\bibitem{kreps2011kafka}
J.~Kreps, N.~Narkhede, J.~Rao \emph{et~al.}, ``Kafka: A distributed messaging
  system for log processing,'' in \emph{Proceedings of the NetDB}, vol.~11,
  2011, pp. 1--7.

\bibitem{pfaff2015design}
B.~Pfaff, J.~Pettit, T.~Koponen, E.~Jackson, A.~Zhou, J.~Rajahalme, J.~Gross,
  A.~Wang, J.~Stringer, P.~Shelar \emph{et~al.}, ``The design and
  implementation of open vswitch,'' in \emph{12th $\{$USENIX$\}$ Symposium on
  Networked Systems Design and Implementation ($\{$NSDI$\}$ 15)}, 2015, pp.
  117--130.

\bibitem{chollet2015keras}
F.~Chollet \emph{et~al.}, ``Keras,'' \url{https://keras.io}, 2015.

\bibitem{abadi2016tensorflow}
M.~Abadi, A.~Agarwal, P.~Barham, E.~Brevdo, Z.~Chen, C.~Citro, G.~S. Corrado,
  A.~Davis, J.~Dean, M.~Devin \emph{et~al.}, ``Tensorflow: Large-scale machine
  learning on heterogeneous distributed systems,'' \emph{arXiv preprint
  arXiv:1603.04467}, 2016.

\bibitem{mahalingam2014virtual}
M.~Mahalingam, D.~G. Dutt, K.~Duda, P.~Agarwal, L.~Kreeger, T.~Sridhar,
  M.~Bursell, and C.~Wright, ``Virtual extensible local area network (vxlan): A
  framework for overlaying virtualized layer 2 networks over layer 3
  networks.'' \emph{RFC}, vol. 7348, pp. 1--22, 2014.

\bibitem{shi2016edge}
W.~Shi, J.~Cao, Q.~Zhang, Y.~Li, and L.~Xu, ``Edge computing: Vision and
  challenges,'' \emph{IEEE internet of things journal}, vol.~3, no.~5, pp.
  637--646, 2016.

\end{thebibliography}

\end{document}